\documentclass[fleqn,usenatbib]{mnras}
\usepackage{tabularx}

\usepackage[T1]{fontenc}
\usepackage{xpatch}
\DeclareRobustCommand{\VAN}[3]{#2}
\let\VANthebibliography\thebibliography
\def\thebibliography{\DeclareRobustCommand{\VAN}[3]{##3}\VANthebibliography}

\usepackage{makecell}
\usepackage{caption}
\usepackage{subcaption}
\usepackage{hyperref}
\usepackage{float}
\usepackage{placeins}
\usepackage{tabularx}
\usepackage{graphicx}	
\usepackage{amsmath}	
\usepackage{booktabs,multirow}
\usepackage{soul}
\usepackage[table]{xcolor}
\usepackage{changepage,threeparttable}
\newcommand{\code}[1]{\texttt{#1}}

\title[Galaxies, haloes and the Cosmic Web]{The relation of galaxies and dark matter haloes to the filamentary cosmic web}

\author[Navdha et al.]{
Navdha$^{1,2}$\thanks{E-mail: 210260035@iitb.ac.in},
Philipp Busch$^{2}$, and 
Simon D. M. White$^{2}$
\\
$^{1}$Indian Institute of Technology Bombay, Mumbai-400076, India\\
$^{2}$Max-Planck-Institut f\"ur Astrophysik, Postfach 1317, D-85741 Garching, Germany\\
}

\date{Accepted XXX. Received YYY; in original form ZZZ}
\pubyear{2024}
\begin{document}
\label{firstpage}
\pagerange{\pageref{firstpage}--\pageref{lastpage}}
\maketitle
\begin{abstract}
We use the Millennium Simulation to study the relation of galaxies and dark matter haloes to the cosmic web. We define the web as the unique, fully connected, percolating object with (unsmoothed) matter density everywhere exceeding 5.25 times the cosmic mean. This object contains 35\% of all cosmic mass but occupies only 0.62\% of all cosmic volume. It contains 26\% of dark matter haloes of mass $10^{11}M_\odot$, rising to 50\% at $10^{12.7}M_\odot$, and to $>90\%$ above $10^{14}M_\odot$. In contrast, it contains 45\% of all galaxies of stellar mass $10^{8.5}M_\odot$,
rising to 50\% at $10^{10}M_\odot$, to 60\% at $10^{11}M_\odot$ and to 90\% at $10^{11.5}M_\odot$. This difference arises because a large fraction of all satellite and backsplash galaxies are part of the cosmic web. Indeed, more than 50\% of web galaxies are satellites for stellar masses below that of the Milky Way, rising to about 70\% below $10^{10}M_\odot$, whereas centrals substantially outnumber satellites in the non-web population at all stellar masses. As a result, web galaxies have systematically lower specific star-formation rates (sSFR's) than non-web galaxies. For the latter, the distributions of stellar mass and sSFR are almost independent of web distance. Furthermore, for both central and satellite galaxies, the sSFR distributions at given stellar mass are very similar in and outside the web, once differences in backsplash fraction are accounted for. For the galaxy formation model considered here, differences between web and non-web galaxy populations are almost entirely due to the difference in halo mass distribution between the two environments.
\end{abstract}

\begin{keywords}
 methods: data analysis -- large-scale structure of Universe

\end{keywords}


\section{Introduction}
As the galaxy population began to be surveyed over 
volumes larger than individual clusters or superclusters, it became
clear that high density regions are linked in an overall
large-scale structure. Thus, \cite{deLapperent1986} noted that the
distribution in their extended CfA survey ``looks
like a slice through the suds in the kitchen sink.. .. galaxies are on
the surfaces of bubble-like structures with diameters of 25 --
50~$h^{-1}$Mpc''. Images from much larger surveys such as 2dF
\citep{Colless2001} and SDSS \citep{Stoughton2002} hardened the
perception that galaxy groups and clusters are joined by a connected
network of filaments and walls which surround large regions of low
galaxy density, so-called ``voids''. In parallel, numerical
simulations of cosmic structure formation became able to cover
relatively large cosmic volumes with sampling
densities much higher than those of observed galaxies, and their
late-time mass distributions showed analogous clusters, filaments and
voids \citep{White1987}. In the current era of large, high-resolution
simulations and high-quality rendering methods, the filamentary
impression conveyed by images of the large-scale mass distribution
is very strong and seems to correpond both qualitatively and
quantitatively to the pattern seen in the galaxies
\citep{Springel2006}. \cite{Bond1996} called this pattern "the cosmic
web", and explained that it can be understood as
reflecting nonlinear gravitational enhancement of a pattern
already present in the initial conditions from
which all structure formed

Over the last 30 years qualitative and quantitative study of these
patterns has been extended using many different characterisations of
the cosmic web. all designed with different scientific goals
and different types of observational or numerical data in mind. Summaries of the philosophy and methodology underlying these various approaches can be found
in the thorough comparison paper of \cite{Libeskind2018} and in the many
individual contributions to the Tallinn IAU Symposium \citep{Weygaert2016}. In the current paper we focus exclusively on
the particular web definition of \citet[][hereafter BW20]{philipp}. Designed for application to high-resolution cosmological
simulations, this assumes the dark matter density profiles of individual galaxy haloes to be adequately
resolved and takes the web to be the unique, spatially
percolating object for which the unsmoothed internal density is
everywhere at least five times the cosmic mean. In practice the
density field and its connectivity are defined by Voronoi tesselation
of the dark matter particle distribution in a straightforward extension of the
group-finding algorithm of \cite{Neyrinck2005}. 

The BW20 scheme has several major advantages over other web definitions for our current purposes. It defines the cosmic web as a material object bounded by an isodensity contour, thus directly analogous to conventional dark matter haloes. It has no adjustable
parameters other than the density threshold, which is determined within a narrow window. Every point and every object in the
simulation volume is either fully part of the web or is a well-defined
distance from its boundary. The large-scale properties of the web depend
only weakly on simulation resolution provided this is high enough to resolve the outer density profiles of the relevant halos. Thus, BW20 found the web to contain about a third of all
cosmic mass within less than a percent of all cosmic volume 
both in the Millennium and in the Millennium-II simulation, despite the
mass resolution of the two differing by a factor of 125. In both, the web is a fully connected network of filaments of typical (but highly variable) width $\sim 1~h^{-1}{\textrm Mpc}$, while the non-web regions also percolate, with the median distance to the nearest web particle being $\sim 7~h^{-1}{\textrm Mpc}$. For any population of objects, it is then straightforward 
to calculate the fraction that are part of the web, and to measure how their properties differ from those of their non-web counterparts; for the latter, one can also measure how their properties depend on distance from the web. The principal goal of this paper is to carry out such an analysis both for dark matter haloes as a function of their mass and for galaxies as a function of their stellar mass, their star formation rate, and their halo environment.

Work with significant similarities to our own has recently been published by \citet{zakharova2023}. This study was also based on the Millennium Simulation but used a different semianalytic galaxy formation model and a different filament-finding algorithm. It concentrated on the level of agreement between filaments defined using the full dark matter distribution and filaments defined using variously selected galaxy samples. Such a study would also be of interest for the galaxy formation model and filament finder used here, but this is not 
the focus of the current paper, which is rather concerned with systematic differences in the properties of haloes and galaxies depending on whether they are or are not part of the cosmic web, as defined from the high-resolution dark matter distribution using the BW20 prescription. 

There is much previous work more distantly related to our own, based on different filament and or web identification algorithms and different kinds of simulated or observed galaxies. \cite{Libeskind2018}
considered twelve different web identification schemes grouped into six different methodological categories.  Comparing their results for a particular simulated dark matter distribution (they did not consider galaxies) led to a 30 author paper occupying 24 journal pages. The BW20 algorithm differs from all of these schemes, and a comparison with even a fraction of them would unbalance our paper and obscure its primary purpose. Furthermore, the high-resolution real-space representation of the dark matter distribution presupposed by the BW20 algorithm is not observationally accessible, so any comparison with real galaxies would require investigation of the relation of the BW20 web to one created from a realistic galaxy catalogue, raising issues concerning the effects of low sampling density, of galaxy biassing, and of redshift space distortions. Although interesting, these issues go beyond our goals for this paper. Because of these complications, we here concentrate exclusively on exploring the relation between haloes, galaxies and the web for one particular web definition and one particular galaxy formation model.

In \S 2 below we briefly describe the dark matter simulation we use and the halo, subhalo and merger-tree structures defined on it (\S 2.1). Next we outline the semianalytic galaxy formation model which provides the present-day galaxy population we analyse (\S 2.2). Its physical parameters were adjusted so that at $z=0$ it reproduces observed galaxy abundances and two-point correlation functions as a function both of stellar mass and of specific star-formation rate. \S 2.3 then describes how BW20 construct the cosmic web and characterise the distribution of web distance for non-web objects. \S 3 presents our results, starting with an image of how the BW20 web structure is reflected in the galaxy distribution. \S 3.1 describes how the web fractions of haloes and galaxies vary with their mass and, for the latter, with halo environment (i.e. central or satellite galaxy). \S3.2 then extends this to examine how halo and galaxy mass functions depend on distance from the web, finding no dependence. \S 3.3  considers whether galaxy star-formation rates depend on web environment. No substantial dependence is found either for central or for satellite galaxies once ``backsplash galaxies'' (i.e. $z=0$ central galaxies which were part of a more massive halo at some time in their past) are removed from the sample of centrals. However, since both satellite and backsplash galaxies are much more common inside the web than outside it, a substantially larger fraction of the overall galaxy population is passive inside the web than outside it. Finally, \S 4 discusses some implications of our analysis and summarises our results.

\section{Data \& Methodology}
\subsection{The Simulation}
In this paper, we will use particle data, merger trees, and galaxy catalogues from the Millennium Simulation (MS; \citealt{ms}). At the time of its completion, the MS was the largest ever cosmological structure formation simulation. It followed the evolution of $10^{10}$ dark matter particles from $z=127$ to $z=0$ throughout a periodic cube of side $500h^{-1}$cMpc in a flat-$\Lambda$CDM cosmology. With a spatial resolution of $5 h^{-1}$ckpc, the MS has poorer resolution but provides better statistics for large-scale structure than its smaller volume companion, the Millennium Simulation II (MSII; \citealt{ms2}). The cosmological parameter choice for both simulations was based on combined analysis of the 2dFGRS
(\citealt{2dfgrs}) and the first year of WMAP data (\citealt{wmap}). Although these parameter values differ significantly from more recent measurements (e.g. \citealt{des}, \citealt{planck}), the cosmic web definition used in our study is based on \citet[][BW20 hereafter]{philipp} who used the MS with its original parameters, and we wish to maintain a close correspondance with their analysis. Moreover, \cite{aw10} showed that the simulation can be scaled straightforwardly to represent clustering in a universe with the currently preferred parameters and the web structure considered in this paper is independent of the overall scale assigned to the simulation. Some of the parameters defining the Millennium Simulation are given in Table \ref{tab:ms_par}, including the total particle
number, $N_{\text{part}}$, the mass of an individual particle, $m_\text{part}$, the side-
length of the simulation’s periodic cubic volume, $L_\text{box}$ and its Plummer-
equivalent gravitational softening, $\epsilon$. The simulation output was stored at 64 output time slices or "snapshots", at each of which friends-of-friends \citep[FOF:][]{fof} group catalogues were computed, grouping together all particles with separation less than 0.2 times the mean particle separation; each FOF group consisting of at least 20 particles was retained. To identify all the bound substructures (subhaloes) within these FOF haloes, an extended version of the \code{SUBFIND} (\citealt{subfind}) algorithm was employed. These subhaloes were then used to build hierarchical merging trees that
describe in detail how structures build up over cosmic time. A subsequent application of semi-analytic models (SAM) of galaxy formation to these subhalo merger trees can be used to construct galaxy catalogues. 

\begin{table}
\centering
\begin{tabularx}{0.525
\columnwidth}{|X|X|}
\hline
 Parameter & Value \\ \hline
$\Omega_{dm}$ & 0.205 \\ \hline
$\Omega_b$ & 0.045 \\ \hline
$\Omega_\Lambda$ & 0.75\\ \hline
$h$ & 0.73 \\ \hline
$\sigma_8$ & 0.9 \\ \hline
$n_s$ & 1 \\ \hline
$N_{\text{part}}$ & $2160^3$ \\ \hline
${m_{\text{part}}}/\left({h^{-1} M_\odot}\right)$ & $8.61 \times 10^8$ \\ \hline
${L_{\text{box}}}/\left({h^{-1} \text{Mpc}}\right)$ & 500 \\ \hline
$\epsilon \left({h^{-1} \text{kpc}}\right)$ & 5 \\ \hline
\end{tabularx}

\caption{Simulation parameters for the MS}

\label{tab:ms_par}
\end{table}

\subsection{Galaxy catalogues}

In this paper, we use the galaxy catalogue built on the MS with its original cosmology by \citet{guo}. Their semi-analytic approach adopts a comprehensive treatment of physical processes involved in galaxy formation and is in general a modified version of previous models, including those of \citet{ms}, \citet{croton},
and \citet{gal_model1}. These include the identification and classification of satellite galaxies within their host haloes, a mass-dependent
model for supernova feedback, the gradual stripping and
disruption of satellite galaxies,  realistic treatments of
the growth of gaseous and stellar disks, a model to calculate bulge and elliptical galaxy sizes, and a reionization model. The distributions of galaxy properties show excellent convergence between the MS and the MSII at stellar masses above $10^{9.5}M_\odot$. 

For the purposes of this paper, it is important to recognise that this scheme does not explicitly model the hydrodynamics of environmental effects on galaxies; there is a simple model for ram-pressure stripping, but it is applied only within the virial radius of dark matter haloes. Hence there may be significant ram-pressure or evaporative effects associated with unvirialised objects like filaments that would be present in a hydrodynamical simulation but are not present here. Tidal effects of the environment on galaxies {\it are} included in the approximation that dark matter and diffuse baryons are similarly affected. Tidal stripping and disruption of the stellar component of galaxies is also included in a simple approximation. 

In this paper, we will use the following properties from the public \citet{guo} galaxy catalogues: galaxy type, which indicates whether the galaxy is at the centre of the most massive subhalo of its FOF group (type~$=0$, a ``central'' galaxy), is at the centre of a less massive subhalo of this group (type~$=1$, a ``satellite'' galaxy), or is a satellite that has lost its subhalo (type~$=2$, an ``orphan'' satellite); snapshot index (we will be mainly concerned with \code{snapnum}$=63$, corresponding to $z=0$); 3D positions; the virial mass $M_{200c}$ of the FOF group that the galaxy is associated with; stellar mass and star formation rate. We also use the galaxy merger trees to identify what are known as ``backsplash'' galaxies \citep{Gill2005}. These are galaxies that are currently type 0 but were type 1 at some time in their past, indicating that they had fallen into a more massive object (such as a galaxy cluster) and later exited again. We find these objects by following back the main progenitor branch of each $z=0$ central galaxy and identifying those which passed through a type 1 phase, at the end of which the virial mass of their halo dropped significantly.

\subsection{The cosmic web}
The cosmic web definition used here is taken from BW20. They identified the cosmic web as the largest object above a suitably chosen density threshold in a density field constructed by Voronoi tessellation of the particle distributions. In practice, they constructed a Tessellation Level Tree (TLT), a hierarchical structure defined on the full unsmoothed particle set of the simulations. Given the density associated to each particle (obtained as its mass divided by the volume of its Voronoi cell) and the list of all the neighbours with which it shares a Voronoi face, the particles are partitioned into disjoint sets (named ``peaks'') each of which is associated through recursive links to its densest neighbour to a specific local density maximum. Spurious peaks due to discreteness noise were suppressed by a persistence filter, requiring the ratio of peak density to limiting density (the density of the highest ``saddle-point'' particle linking a peak it to its higher density ``parent''), $r \ge 10$. However, when the TLT is used to define ``friends-of-friends'' groups of particles above a chosen density threshold, the particle set identified with each group is independent of this persistence filtering. 

As the density threshold is varied, the group population shows
a pronounced and unambiguous percolation transition at $\rho_\text{thresh} \sim 7\left < \rho \right >$\footnote{This number depends on the number density of sampling points (\citealt{sd_dep}) - for example, it is $9\left<\rho\right>$ for the MSII which has 125 times larger particle density than the MS.} and for all threshold densities below this, the largest object has much greater mass than the second largest, is fully spatially connected in all three spatial dimensions and can be defined as the cosmic web. We use a threshold density of $5.25\left < \rho \right >$ in this paper. For this choice the cosmic web contains 35\% of the mass and occupies 0.62\% of the volume of the MS, giving it an average density about 60 times the cosmic mean. Note that at every time, each galaxy is associated with a specific simulation particle, inherits its position and velocity, and is thus unambiguously either part of the web or not. Furthermore, individual galaxy haloes are either fully part of the web or entirely disjoint from it. Note also that although BW20 demonstrated that their web has primarily filamentary morphology\footnote{See the video at https://pbusch.net/tlt/perc\_movie.mp4 and the associated discussion in \S 4.3 of their paper.}, we do not attempt to differentiate between filaments and haloes, groups, clusters or walls in this paper.

If we consider two sets of points distributed throughout the simulation volume, we can characterise the distribution of B around A by histogramming the distance $d$ from each point in B to its nearest neighbour in A. We will use the resulting cumulative probability distributions $F(<d)$ below to study how galaxies, galaxy haloes and dark matter as a whole are distributed around the cosmic web. In these applications A will be the set of dark matter particles belonging to the web, while B will be either a subpopulation of galaxies or galaxy haloes, or a set of randomly chosen dark matter particles or uniformly distributed points in space. The fraction of objects which are part of the cosmic web is then defined by $d\approx 0$, while the distribution at $d>0$ indicates how broadly the remaining objects are distributed around the web.\footnote{In practice, the distributions $F(<d)$ can be calculated efficiently even for very large sets A and B using the algorithm
\texttt{KDTree} as implemented, for example, in the \texttt{Python} package \code{sklearn.neighbors.KDTree}. } A version of this approach was already used in BW20 for the case where the set B was a uniform $1024^3$ grid filling the simulation volume. They referred to the resulting distance distributions as the Euclidean Distance Transform (EDT) of the cosmic web.

\begin{figure*}
	\includegraphics[width=150mm]{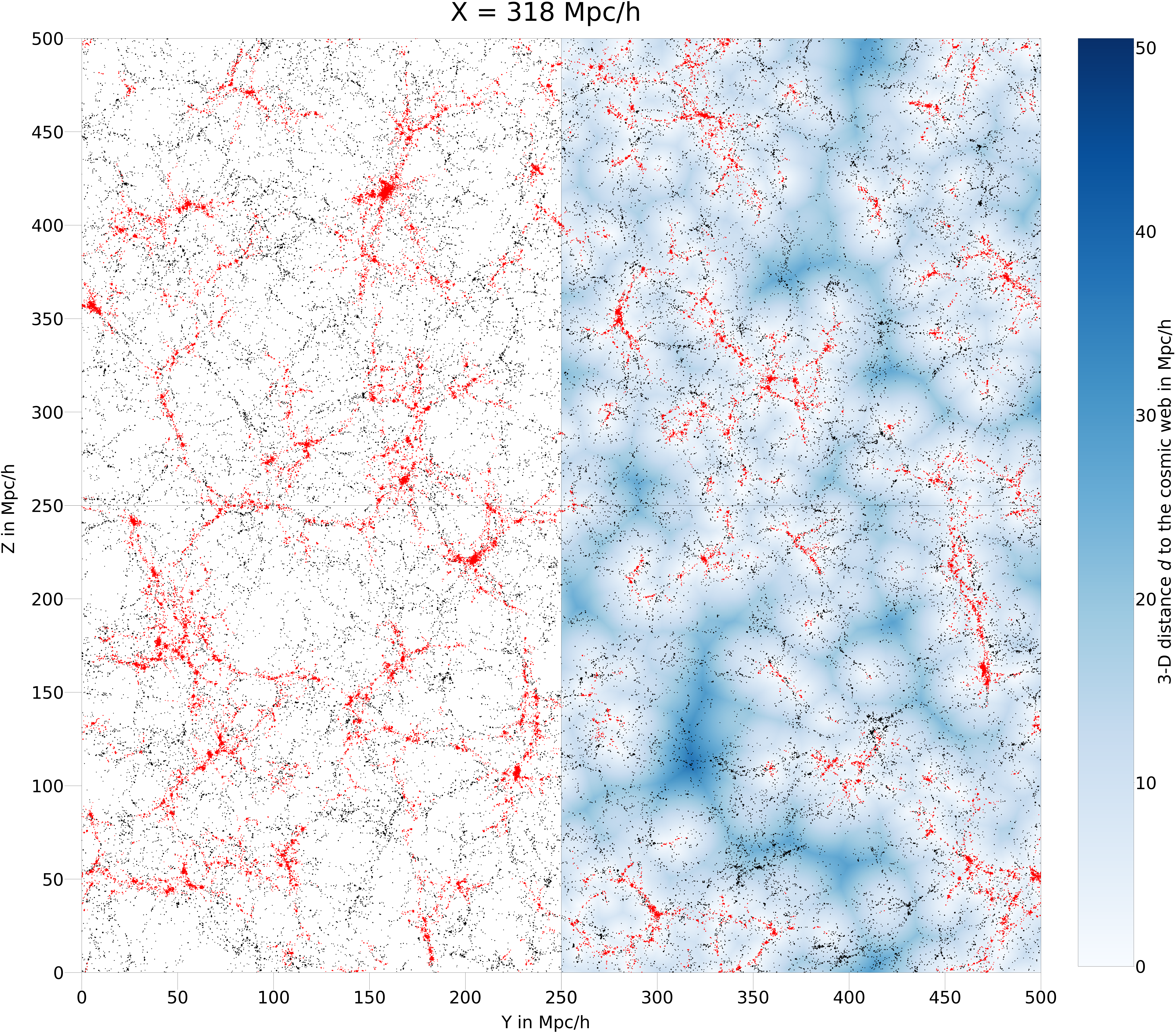}
    \caption{A $20h^{-1}$Mpc thick slice along the X-coordinate showing all galaxies with stellar mass $M_* > 10^{9.5}M_\odot$. Those which are part of the cosmic web are coloured red, the rest black. The blue shading in the right half of the plot indicates the minimum distance $d$ from each point in the midplane to a particle in the web (the so-called EDT). BW20 chose this slice to include the global maximum of the EDT field (in the bottom right quadrant) and thus the largest void in the MS.} 
    \label{fig:full}
\end{figure*}
\section{Results}
Fig.~\ref{fig:full} shows the distribution of $z=0$ galaxies with stellar mass greater than $10^{9.5}M_\odot$ in a $20 h^{-1}$Mpc thick slice of the MS centred at $X = 318h^{-1}$Mpc. Galaxies are coloured red if they are part of the cosmic web, as defined in the last section, black if they are not. The midplane of this slice corresponds to the plane shown in Fig.~8 of BW20, which they chose to pass through the centre of the largest sphere containing no web particle (the biggest ``void''). The blue shading in the right-hand side of this plot corresponds to that in their figure and indicates the 3D distance from each point in the midplane to the nearest web particle (what BW20 call the EDT). The filamentary nature of the galaxy distribution is clear in this plot, not only in the red points, which we here consider to be part of the cosmic web, but also in the black points which we consider to be outside it. It is noticeable, however, that all the relatively massive clusters of galaxies are red. We now move on to more quantitative analysis of the relation of galaxies and their haloes to this cosmic web.

\subsection{The mass dependence of web fractions}

\begin{figure*}
	\centering
	{\label{figur:1.1}\includegraphics[width=85mm,height=70mm]{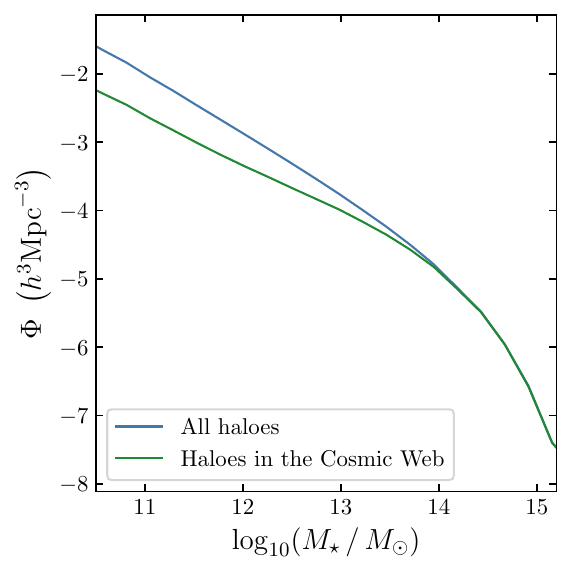}}
{\label{figur:1.2}\includegraphics[width=85mm,height=70mm]{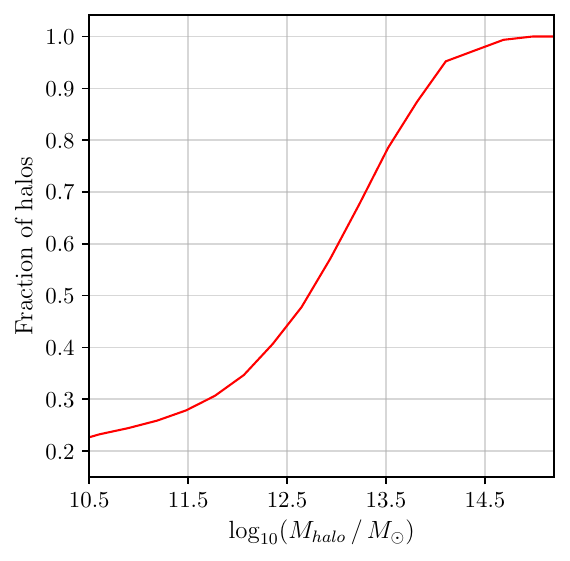}}
\\
{\label{figur:2.1}\includegraphics[width=85mm,height=70mm]{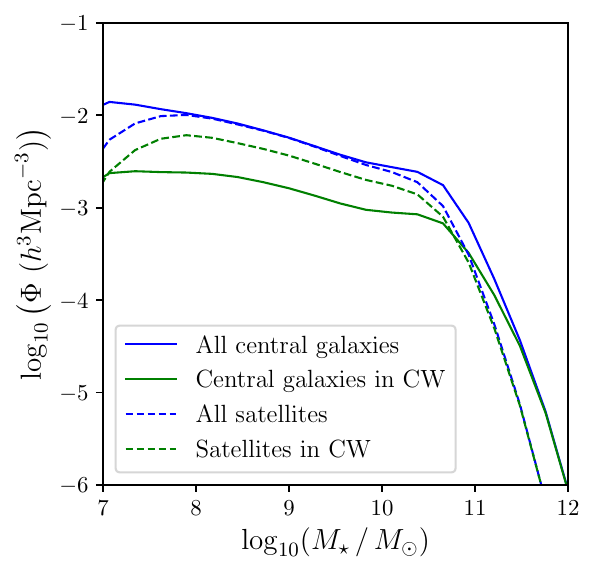}}
{\label{figur:2.2}\includegraphics[width=85mm,height=70mm]{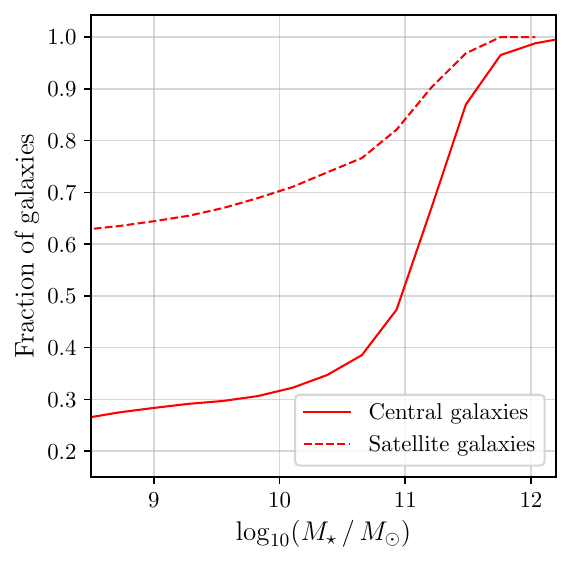}}
\\
{\label{figur:3.1}\includegraphics[width=85mm,height=70mm]{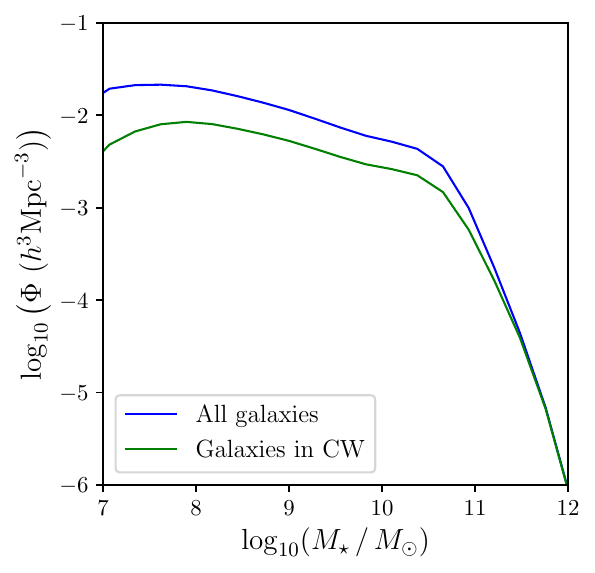}}
{\label{figur:3.2}\includegraphics[width=85mm,height=70mm]{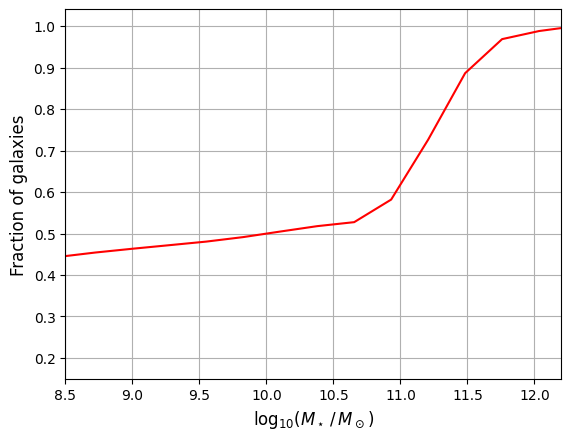}}
    \caption{Mass functions and web fractions for dark matter haloes (1st row), central and satellite  galaxies (2nd row), and all galaxies (3rd row). In the mass function panels on the left, the overall population is shown in blue and the subpopulation belonging to the web in green. The sharp drop in the stellar mass functions at low mass reflects the limited mass resolution of the MS; the MS agrees reasonably well with the higher resolution MSII for $M_* > 10^{8.5}M_\odot$ and almost perfectly for $M_* > 10^{9.5}M_\odot$ \citep{guo}. We therefore limit the web fraction plots on the right to stellar masses above the former limit. The web fractions increase with mass in all three cases, with the main change occurring in a substantially narrower mass range for galaxies than for haloes. In addition, because satellite galaxies are predominantly in the web, almost half of all galaxies are still in the web for $M_* < 10^{10}M_\odot$, whereas for central galaxies this fraction is less than a third.}
    \label{fig:full2}
\end{figure*}

As noted earlier, in the MS 35\% of all dark matter particles and 0.62\% of randomly chosen points in space are part of the cosmic web as we defined it in \S 2.3. For other populations of objects, this fraction varies substantially. A dark matter halo is part of the web if its central particle is a web particle. As shown in Fig.~\ref{fig:full2}, the probability of this happening is a strong function of halo mass. The top left panel compares the overall abundance of haloes as a function of $M_{200c}$ (the halo mass function\footnote{We characterise haloes by $M_{200c}$, the mass within the largest sphere centred on their potential minimum for which the enclosed mean density is at least 200 times the critical value.}) to the abundance of the subset which are part of the web; the top right panel shows the ratio of these two curves. The web fraction varies from 20\% to 100\% over the mass range considered. Haloes are less likely than a random dark matter particle to be part of the web for $M_{200c} < 10^{12.3}M_\odot$ and more likely in the opposite case. Interestingly, the transition occurs at a mass typical of the haloes of isolated bright spirals like the Milky Way and M31.

There is a relatively tight relation between the total mass of a halo and the stellar mass of its central galaxy which has a pronounced feature at about this same halo mass (e.g. Fig.~9 of \citealt{guo}). At lower mass, the ratio $M_*/M_{200c}$ increases quite rapidly with halo mass, while at higher mass it decreases again. As a result, the fraction of central galaxies which are part of the web is slowly varying and relatively small at stellar masses less than that of the Milky Way, 
but increases rapidly at higher stellar mass (see the solid curves in the two central panels of Fig.~\ref{fig:full2}); the main increase occurs over a substantially narrower mass range than in the halo case. Isolated galaxies less massive than the Milky Way are less likely to be part of the web than random dark matter particles, but the web fraction increases rapidly at higher stellar mass, reaching 100\% for the most massive objects.

While most galaxies are centrals at high stellar mass, the satellite fraction increases steadily as stellar mass decreases, and for stellar masses between $10^8$ and $10^{10}M_\odot$ satellites and centrals are almost equally abundant (compare the dashed and solid blue curves in the middle left panel of Fig.~\ref{fig:full2}). The middle right panel of Fig.~\ref{fig:full2} shows that web fractions of the two kinds of galaxy are very different at all stellar masses, however.
This is because satellites almost always live in substantially more massive haloes than centrals of the same stellar mass. As a result, for stellar masses below that of the Milky Way the majority of web galaxies are satellites, and for $M_*<10^{10}M_\odot$ satellites outnumber centrals in the web  by more than a factor of two. Outside the web, the galaxy population is dominated by central galaxies at all stellar masses. When galaxies of all types are considered together, the web fraction is substantially increased at low stellar mass. This effect is seen in the lowest two panels of Fig.~\ref{fig:full2}.
Most galaxies are part of the web for all stellar masses above $10^{10}M_\odot$ and even for $M_* = 10^{8.5}M_\odot$ the web fraction is 45\%.
\subsection{Dependence on distance from the Web}

We use cumulative distance distributions $F(<d)$ as defined in \S~2,3 to study the mass dependence of the distribution of galaxies and galaxy haloes relative to the cosmic web, and to compare these distributions with those of random points in space and of random dark matter particles. 

\begin{figure*}
	\centering
	{\label{fig:3.1}\includegraphics[width=85mm]{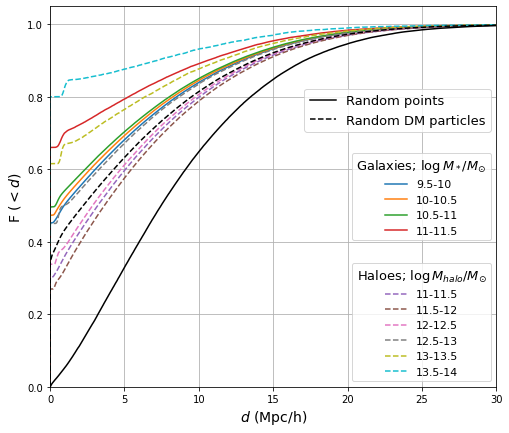}}
{\label{fig:3.2}\includegraphics[width=85mm]{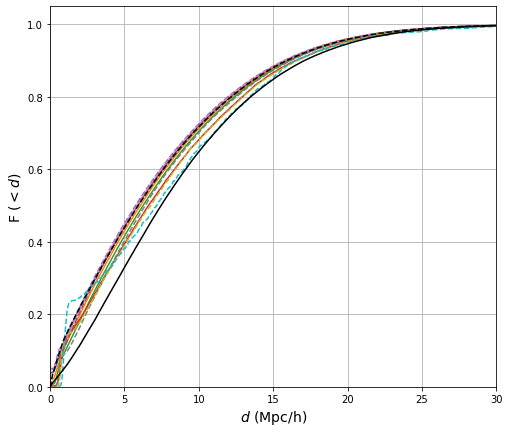}}
    \caption{Normalized cumulative distributions of distance to the cosmic web for galaxies and haloes as a function of their mass, as well as for random dark matter particles and for random points in space. In the left panel the curves are for all objects in each class, so that their value at $d\rightarrow 0$ corresponds to the fraction which are in the web (see also Fig.~\ref{fig:full2}). The right panel is similar but considers only the subpopulation of each class which is not part of the web, so that $F(d\rightarrow 0) = 0$ for all curves. Haloes and galaxies are very similarly distributed around the web to the dark matter itself at all but the highest masses, for which the distributions become broader on large scales, approaching that of random points in space. Because of the large volume of the MS, counting noise is negligible in all of these curves.}
    \label{fig:full3}
    
\end{figure*}

The left panel of Fig. \ref{fig:full3} shows these distributions for all objects in specific mass ranges, as indicated in the figure legend. At $d\sim 0$, the curves in this panel start from the mass fractions shown in Fig.~\ref{fig:full2}. Because haloes have a finite size, scaling with the cube root of their mass, objects which are not part of the web are constrained to be at least one halo radius away from it. This gives rise to the features seen on small scale in these plots, particularly for the most massive objects. For example, 80\% of the most massive haloes are part of the web. There are then no objects with $0 < d < 700 ~\textrm{kpc/h}$, but a sharp rise to include another 3\% within about 1.0 Mpc/h. These are objects that connect to the web over a bridge whose minimum (saddle-point) density is somewhat below our threshold of $5.25\langle{\rho}\rangle$. The $F(<d)$ distribution of random dark matter particles is almost identical to that of haloes with mass $12.0< \log_{10}(M_{\textrm{halo}}/M_\odot)< 12.5$, whereas the distribution for random points appears broader.

In the right panel of Fig.~\ref{fig:full3} we show the same data, except we now restrict to the subpopulation of each class of object which is {\it not} part of the web. Thus, these distributions all go to zero as $d\rightarrow 0$. The most striking thing about these curves is their similarity. The solid black curve for random points in space is the exact equivalent of the EDT curves shown by BW20. The dashed black curve for random dark matter particles is steeper than this at small $d$, indicating some concentration of dark matter towards the web, but this effect
is less strong than might have been anticipated. For example, averaged over the range $0<d<1.0~\textrm{Mpc/h}$, the mean density of dark matter is only 1.67 times the cosmic mean, while for $0<d<5.0~\textrm{Mpc/h}$ this factor is just 0.84. The median $d$-value for non-web dark matter is 5.9~Mpc/h whereas for random points in space it is 7.5~Mpc/h.

Except at the highest masses, the $F(<d)$ curves for galaxies and haloes are all very close to that for dark matter particles in the right panel of Fig.~\ref{fig:full3}. At the highest masses the exclusion and proximity effects discussed in the last paragraph are visible on scales below a few Mpc, while on relatively large scale the curves shift away from that for random dark matter particles towards that for random points in space. Thus, while the great majority of non-web galaxies are distributed around the web in the same way as the dark matter, independent of their stellar mass, the most massive are somewhat more broadly distributed on large scale, consistent with them uniformly filling the volume not occupied by the web.

\subsection{Specific star formation rate dependences}

The previous sections have shown that while haloes and galaxies of all masses are present both in the web and outside it, the web fraction of both types of object increases with their mass. Outside of the web, however, the mass functions both of haloes and of galaxies are almost independent of distance to the web.\footnote{If the $F(<d)$ curves in the right panel of Fig.~\ref{fig:full3} coincided exactly, the halo and galaxy mass functions would be fully independent of $d$. The slight rightward shift of the curves for the most massive objects means, however, that the relative abundance of these objects is somewhat enhanced far from the web.} In this section we investigate whether the specific star formation rate (sSFR, defined as $d\ln M_*/dt$) of galaxies of given stellar mass depends on whether they are part of the web or not, and in the latter case whether there is a dependence on distance from the web. It is well known, of course, that the sSFR of galaxies of given stellar mass depends strongly on their halo environment, In particular, satellite galaxies inhabit more massive haloes and have lower sSFR than central galaxies of the same stellar mass. In the \cite{guo} galaxy formation model used here, this effect is strong (and, indeed, is rather stronger than observed, see \cite{Hirschmann2014}) so the bias of massive haloes towards the web can be expected to induce a bias at fixed stellar mass in favour of satellites, and hence lower sSFR, in the web galaxy population.

\begin{figure*}
	\centering
	{\label{fig:4.1}\includegraphics[width=56mm]{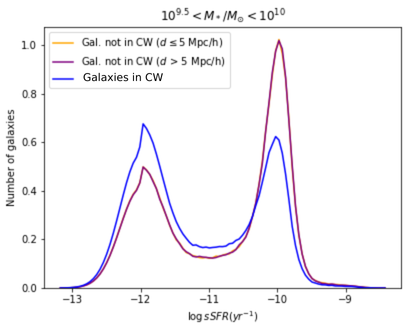}}
{\label{fig:4.2}\includegraphics[width=56mm]{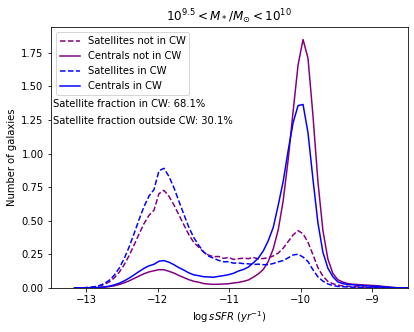}}
{\label{fig:4.3}\includegraphics[width=56mm]{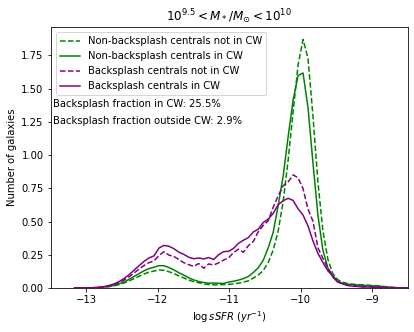}}
\\
{\label{fig:4.1}\includegraphics[width=56mm]{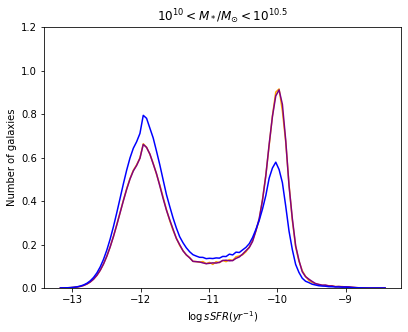}}
{\label{fig:4.2}\includegraphics[width=56mm]{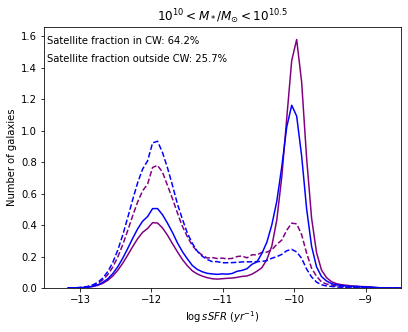}}
{\label{fig:4.3}\includegraphics[width=56mm]{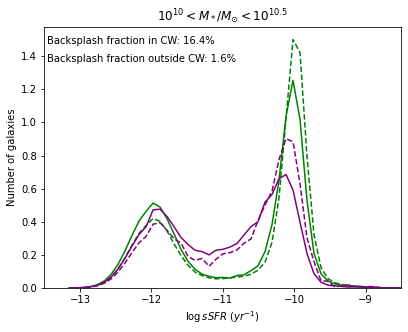}}
\\
{\label{fig:4.1}\includegraphics[width=56mm]{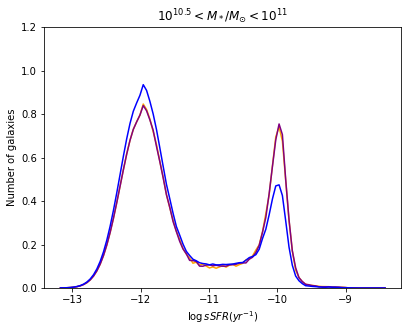}}
{\label{fig:4.2}\includegraphics[width=56mm]{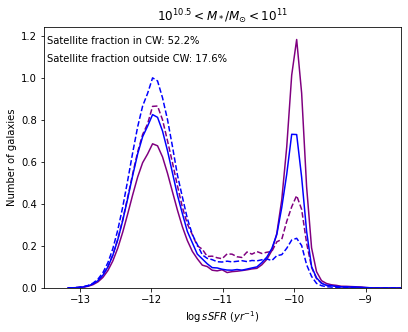}}
{\label{fig:4.3}\includegraphics[width=56mm]{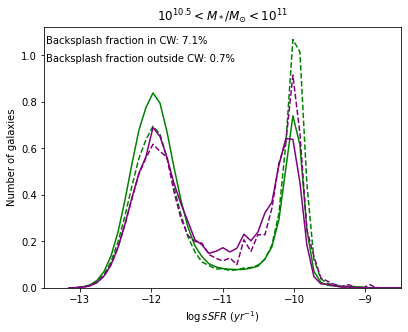}}
\\
{\label{fig:4.1}\includegraphics[width=56mm]{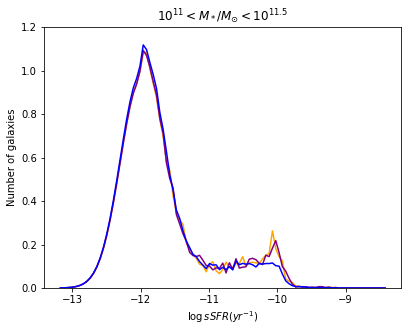}}
{\label{fig:4.2}\includegraphics[width=56mm]{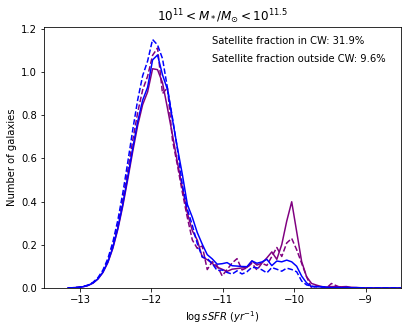}}
{\label{fig:4.3}\includegraphics[width=56mm]{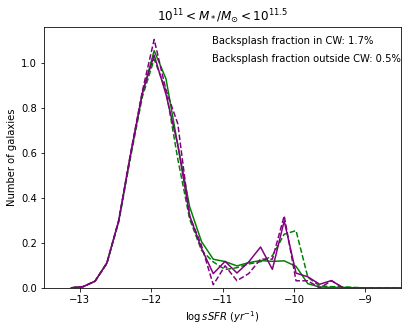}}
    \caption{Normalized sSFR distributions in mass ranges indicated by a legend above each panel for galaxies in and outside the cosmic web (CW). The left panels show results for all galaxies, splitting non-web objects into two groups according to distance from the web. The central panels separate galaxies into those which are central within their halo and those which are satellites, with a legend indicating the fractions of galaxies in and outside the web which are satellites. The right panels split central galaxies into those which were once satellites within a more massive halo (backsplash galaxies) and those which were not, a legend here indicating the fractions of central galaxies in and outside the web which are backsplash. All plots show two peaks - most galaxies are either passive (the peak at $\sim -12$ or star-forming (the peak at $\sim -10$). The shape (but not the amplitude) of the former peak is artificial (see the text). Counting noise is very small in most of these curves, but is visible as small-scale bin-to-bin variations in the curves for the most massive galaxies and for backsplash galaxies.}
    \label{fig:full4}
    
\end{figure*}

The four rows of panels in Fig.~\ref{fig:full4} show normalised sSFR distributions for galaxies in four different stellar mass ranges, each 0.5~dex wide. To ease interpretation, passive galaxies with $\log_{10} \textrm{sSFR}/(\textrm{yr})^{-1} < -12$ are represented in all panels by a suitably normalised gaussian centred at $-12$ and with standard deviation 0.3. As a result, almost all the sSFR distributions show two distinct and well separated peaks, and while the shape and position of the passive peak are set arbitrarily, those of the peak of actively star-forming galaxies at $\log_{10} \textrm{sSFR}/(\textrm{yr})^{-1} \sim -10$ are a direct consequence of the galaxy formation model. The relative amplitude of the two peaks, the position and shape
of the star-forming peak, and the height of the ``bridge'' between the two peaks thus all contain information about the relation between the star-formation properties of galaxies and their web environment.
 \begin{figure}
    \centering
    \includegraphics[width=\columnwidth]{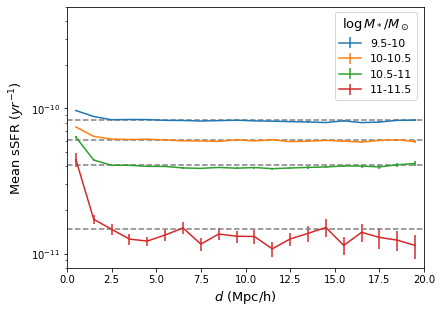}
    \caption{Mean value of sSFR along with the standard error in the mean for the non-web galaxy population plotted as a function of distance to the cosmic web, split into four mass ranges. Values of the sSFR averaged over all distances for each mass range are shown as grey horizontal lines. The increased sSFR below 2~Mpc is due primarily to the reduced satellite fraction resulting from the requirement that massive haloes be at least one halo ``radius'' from the web to avoid being joined to it. }
    \label{fig:5}
\end{figure}

The left panels in Fig.~\ref{fig:full4} show sSFR distributions for galaxies which are part of the web (the blue curves) for galaxies with $0 < d < 5~\textrm{Mpc/h}$ (the orange curves) and for galaxies with $d \geq 5~\textrm{Mpc/h}$ (the purple curves). In fact, the orange and purple curves are extremely close in all four mass ranges and are only separately visible at a few points where counting noise leads to a slight separation. Thus, for the \cite{guo} galaxy formation model, the masses and star formation rates of non-web galaxies are almost independent of web distance (we show this explicitly for the mean sSFR in Fig.\ref{fig:5}). On the other hand, at all stellar masses, the sSFR distributions of web galaxies are different from those of non-web galaxies. Their passive fraction is higher, and their star-forming peaks are lower, with greater asymmetry in the form of a heavier tail towards low sSFR.

The middle panels in Fig.~\ref{fig:full4} explore the extent to which these differences are driven by different proportions of satellite and central galaxies. In these panels the sSFR distributions for web galaxies are shown in blue and those for non-web galaxies in purple. In each case the population is split into central galaxies (the solid curves) and satellite galaxies (the dashed curves). The satellite fractions of the web and non-web galaxy populations in each mass range are indicated by a legend in the relevant panel. These fractions decrease with stellar mass and are always larger for web galaxies than for non-web galaxies. At all stellar masses, satellites have a larger passive fraction, a less narrowly defined star-forming peak and substantially more objects with $\log_{10}\textrm{sSFR}/(\textrm{yr})^{-1} \sim -11$ than central galaxies. For both types the difference in sSFR distribution between web and non-web galaxies is quite small. Thus the differences between web and non-web galaxies seen in the left panels of Fig.~\ref{fig:full4} are driven primarily by the larger satellite fraction in the web.

There is still, however, a significant difference in passive fraction between those central galaxies which are in the web and those which are not (except at the highest stellar mass where essentially all galaxies are passive). We explore the origin of this difference in the right panels of Fig.~\ref{fig:full4}, where the $z=0$ central galaxy populations of the middle panels are split into those which have always been central galaxies and those which were once a satellite of some more massive object. The latter are often referred to as "backsplash galaxies" \citep{Gill2005} because they typically fell through a cluster some time in the past, but then exited again and are currently outside the cluster. In these right panels the sSFR distributions of central galaxies are shown in blue for web objects and in purple for non-web objects, but now solid lines refer to those galaxies which have always been centrals, and dashed lines to those which are backsplash galaxies. The legends in the panels indicate the fraction of each population which is backsplash. At all masses the backsplash fraction is about ten times larger in the web than outside it and indeed, for the lowest mass bin shown, about a quarter of all web centrals are backsplash. The sSFR distribution of backsplash galaxies is significantly different from that of the non-backsplash centrals, with a larger passive fraction, a lower star-forming peak, and many more objects with intermediate sSFR values. In both cases, however, there is little difference between the web and non-web populations. We conclude, therefore, that most of the difference between the sSFR distributions of web and non-web central galaxies seen in the middle panels of Fig.~\ref{fig:full4} is due to the much larger fraction of backsplash galaxies in the web population.

\begin{figure}
    \centering
    \includegraphics[width=\columnwidth]{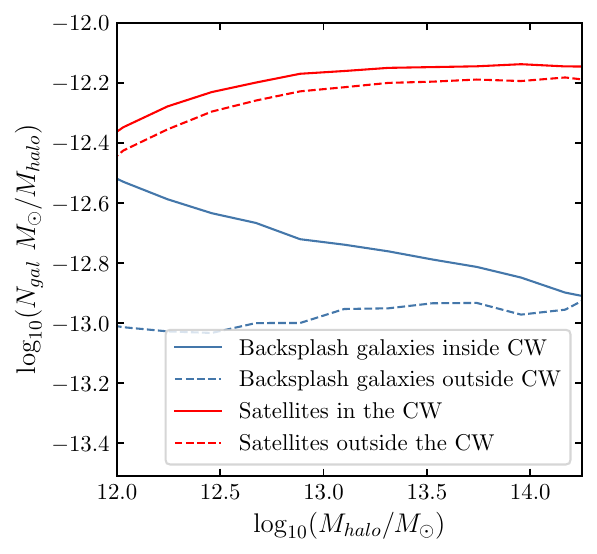}
    \caption{Mean number per unit halo mass of satellite (red) and backsplash (blue) galaxies with stellar mass exceeding $10^{9.5}M_\odot$, plotted as a function of the associated present-day halo mass. The solid and dotted curves correspond to objects in and outside the web, respectively.}
    \label{fig:6}
\end{figure}

Most of the differences discussed
in this section reflect the bias of web haloes towards high mass (Fig.~\ref{fig:full2}), which produces a larger satellite fraction in the web. An additional contributing factor might be that web haloes have systematically more satellites and associated backsplash galaxies than non-web haloes of similar mass. We examine this possibility in Fig.~\ref{fig:6}, which shows the number of satellite and backsplash galaxies more massive than $10^{9.5}M_\odot$ per unit halo mass as a function of the mass of the associated present-day halo, splitting the results according to whether the halo is part of the web or not. Interestingly, these curves show only a weak variation with halo mass over the range plotted. Web haloes do indeed have more satellite galaxies than non-web haloes, but the effect is modest, a 10 to 20\% excess with only a weak dependence on halo mass. A larger effect is seen for backsplash galaxies, ranging from almost a factor of three at $10^{12}M_\odot$ to about 30\% at  $10^{14}M_\odot$. Since on average the number of backsplash galaxies associated with a halo is about an order of magnitude smaller than the number of satellites,  these effects together account for only a small fraction of the stellar population shifts seen in Fig.~\ref{fig:full4}.

\section{Discussion and Conclusions}

In this paper, we have used galaxy catalogues and particle data from the Millennium Simulation to study the dependence of galaxy and halo properties on their cosmic environment, in particular, on their membership in, or distance from the cosmic web. We have chosen to follow \citet[][BW20]{philipp}, who identify the cosmic web as the unique percolating object with $\rho > 5.25\langle\rho\rangle$ in a density field constructed by Voronoi tessellation of the full dark matter particle set. As noted in the introduction, this definition is particularly convenient for our purposes because the resulting structure is a material object directly analogous to standard dark matter haloes and depends only weakly on simulation resolution provided galaxy haloes are resolved, Furthermore it has just one tunable parameter, the threshold density, $\rho_{\rm th}$. Our results are, of course, quantitatively dependent on BW20's choice, $\rho_{\rm th}/\langle\rho\rangle = 5.25$. Their Figs 6 and 7 show that as this ratio is reduced, the fraction of non-web haloes drops precipitously. For example, the number of non-web haloes of Milky Way mass is about three orders of magnitude smaller for $\rho_{\rm th}/\langle\rho\rangle = 1$. Almost all of the apparent structures in our Fig.~\ref{fig:full} would then be red, even though the web still occupies only $\sim 6\%$ of all cosmic volume. The qualitative conclusions we give below would still be valid in this case, however.

With the BW20 definition every halo, galaxy or dark matter particle is either part of the web or not. We have characterised the distribution
of various populations of haloes and galaxies relative to the web using histograms of the distance $d$ from each object to the nearest web particle. These can be calculated efficiently, even for very large populations, using the KDTree algorithm. Objects lying in the web are then those with $d \approx 0$. 
The main results obtained from this analysis for $\rho_{\rm th}/\langle\rho\rangle = 5.25$ may be summarized as follows:
\begin{enumerate}
    \item The web fractions of both haloes and galaxies increase with mass. Only 26\% of all haloes of mass $10^{11}M_\odot$ are part of the web, but this number rises to 50\% at $10^{12.7}M_\odot$ and to
    $>90\%$ at $10^{14}M_\odot$. For comparison, the web contains 35\% of all dark matter. In contrast, 45\% of all galaxies of stellar mass $10^{8.5}M_\odot$ are part of the web, rising to 50\% at $10^{10}M_\odot$, to 60\% at $10^{11}M_\odot$, and to 90\% at $10^{11.5}M_\odot$. 
    \item The relatively high fraction of low-mass galaxies which are part of the web is a consequence of many of them being satellites. Indeed, the web galaxy population is dominated by satellites at all stellar masses below that of the Milky Way; for $M_*<10^{10}M_\odot$ web satellites outnumber web centrals by a factor of about two. In contrast, outside the web, centrals substantially outnumber satellites at all stellar masses.
    \item Outside the web, the mass distributions both of haloes and of galaxies are almost independent of distance to the web. Except for the highest masses, the distance distribution does not depend significantly on mass and is very similar to that of randomly selected dark matter particles. At the highest masses, the distance distributions of both haloes and galaxies are somewhat broader, resembling that of randomly distributed points in space.
    \item Galaxies of given stellar mass have systematically lower star formation rates in the web than outside it. This is primarily because the satellite and backsplash fractions are larger, and star formation is suppressed in both types of galaxy. The star formation rate distribution for non-backsplash central galaxies is similar inside and outside the web, and the same is true both for satellite galaxies and for backsplash galaxies, but satellite fractions are two to three times larger and backsplash fractions are ten times larger in the web. This is primarily caused by the bias of the web halo population towards high halo mass; web haloes have only slightly more satellites and modestly more backsplash galaxies than non-web haloes of the same mass,
    \item For non-web galaxies of given stellar mass, the distribution of star formation rate is independent of distance from the web.  
\end{enumerate}
Conclusions analogous to these have, of course, been found previously for other definitions of the cosmic web and for different populations of simulated or observed galaxies. The web definition most closely related to BW20 is that of the DisPerSE algorithm which is also based on a density field estimated using Voronoi-Delaunay tesselation of a set of structure tracers \citep{Sousbie2011a,Sousbie2011b}. Indeed, if DisPerSE is applied to the dark matter particle distribution of a high-resolution cosmological simulation, the largest connected part of the resulting filamentary web, after deletion of all elements below threshold, forms the ``skeleton'' of the BW20 web.\footnote{Indeed, at the resolution of the Millennium Simulation, DisPerSE would, for persistence values $r\sim 10$ sufficient to remove spurious noise peaks, identify essentially {\it all} haloes containing galaxies like those considered in this paper as peaks, and hence as nodes of its network; {\it no} galaxies would lie more than a halo radius away from these nodes.} The main differences between the two webs are that the DisPerSE connects {\it all} peaks (and hence all haloes and galaxies) resolved by the underlying simulation and consists of line segments (filaments), whereas BW20 includes only elements which are connected by regions above threshold and is a material object bounded by an equidensity contour with finite mass and volume fractions.

\cite{Laigle2018} compared galaxies from a hydrodynamical simulation to the DisPerSE web defined from its dark matter distribution. Their results for the web-distance distribution of galaxies as a function of mass and star-formation activity are very similar to those summarised above, once the fact that the BW20 web extends out to a variable distance (typically around 1 Mpc) from its skeleton is taken into account.  \cite{zakharova2023} used the Millennium Simulation and a very similar semi-analytic galaxy formation model to this paper in order to compare DisPerSE webs constructed from dark matter particles\footnote{They took a very high persistence value $r=10^4$ which implies that only massive haloes corresponding to galaxy groups are retained as peaks.} and from galaxies, but they concentrated on the relation between the two webs, rather than on the systematic variation of galaxy properties with web distance. 

Most studies looking at how galaxy properties vary with web distance have preferred to use the web defined by the galaxies themselves. Examples based on DisPerSE include \cite{Galarraga2020} and \cite{Kuchner2022} for simulated galaxies, and \cite{Kraljic2018} and \cite{Okane2024} for observed galaxies. Proper comparison of our results with this work would require detailed analysis of the relation between BW20 webs for the dark matter and for the much more sparsely sampled galaxies. In addition, for observational samples it is also necessary to understand the effects of redshift-space distortion and/or of photometric distance uncertainties. While these are all interesting issues they go far beyond the goals of this paper. The situation is even more complicated for other filament or web finders \citep[e.g.][]{Cautun2013,Tempel2016} which identify quite different structures than DisPerSE, even for the same underlying simulation \citep{Libeskind2018}.

The semianalytic galaxy formation model we have used (from \citealt{guo}) assumes that the galaxy population in a halo is determined entirely by the halo's mass assembly history (specifically, by its merger tree) and is independent of accretion morphology. If assembly history at  given halo mass were independent of larger scale environment, as is the case in simple excursion set models for the growth of structure \citep[e.g.][]{White1996}, the above results would follow from the fact that the halo mass distribution is biased high in the web but is independent of position outside it. However, the assembly bias phenomenon demonstrates a measurable violation of such environmental independence \citep{Gao2005} as does the greater abundance of backsplash galaxies that we find in the web, even for haloes of similar mass.  Both effects reflect a dependence on environment density, which in turn correlates with web location. The model of \citet{guo} is not only qualitatively consistent with the filamentary appearance of the observed galaxy distribution (see also Fig.~\ref{fig:full}) but also qualitatively consistent with observed galaxy autocorrelation functions as a function of stellar mass and colour. Thus, these clustering properties provide no direct evidence for an influence of web morphology on galaxy formation. This influence may be more convincingly demonstrated by looking for correlations between local web morphology and the shapes \citep{Binngeli1982,Delgado2023} or spins \citep{Navarro2004,Codis2012} of galaxies. It will be interesting to repeat the analysis of this paper for high resolution hydrodynamical simulations of cosmological galaxy formation, since the alignment of intergalactic shocks with filaments may lead to more pronounced effects on galaxy morphology \citep{Benitez2013} .

\section*{Acknowledgements}
We thank the anonymous referee for a constructive and insightful report that helped to improve
  the paper.
\section*{Data Availability}
This work used publicly available halo and galaxy data from the Millennium simulation available in the form of SQL-queryable database at \href{https://wwwmpa.mpa-garching.mpg.de/millennium/\#DATABASE\_ACCESS}{https://wwwmpa.mpa-garching.mpg.de/millennium/\#DATABASE\_ACCESS}. Additional data presented in this article will be shared upon reasonable request to the corresponding author.

\bibliographystyle{mnras}
\bibliography{main} 



\bsp	
\label{lastpage}
\end{document}